\documentclass[runningheads]{llncs}
\usepackage{graphicx}
\usepackage{amsmath}
\usepackage{siunitx}
\usepackage{amsfonts}
\usepackage{amssymb}
\usepackage{mathtools}
\usepackage{bm}
\usepackage{accents}
\usepackage{subcaption}
\usepackage{microtype}
\usepackage{booktabs,makecell}
\usepackage[dvipsnames]{xcolor}
\usepackage{color, colortbl}
\definecolor{Gray}{gray}{0.90}
\usepackage{float}

\usepackage{hyperref}

\usepackage[capitalise,nameinlink]{cleveref}
\begin{document}
\title{GAMMA-PD: Graph-based Analysis of Multi-Modal Motor Impairment Assessments in Parkinson's Disease}

\titlerunning{GAMMA-PD}

\author{Favour Nerrise\inst{1}\orcidID{0000-0002-1959-5302} \and Alice Louise Heiman\inst{2} \and Ehsan Adeli\thanks{Corresponding author.}\inst{3}\orcidID{0000-0002-0579-7763}} 
\authorrunning{F. Nerrise et al.}
\institute{Department of Electrical Engineering, Stanford University, Stanford, CA, USA \and 
Department of Computer Science, Stanford University, Stanford, CA, USA \and 
Dept. of Psychiatry \& Behavioral Sciences, Stanford University, Stanford, CA, USA \\ 
\email{\{fnerrise,eadeli\}@stanford.edu}\\
}
\maketitle              
\begin{abstract}
The rapid advancement of medical technology has led to an exponential increase in multi-modal medical data, including imaging, genomics, and electronic health records (EHRs). Graph neural networks (GNNs) have been widely used to represent this data due to their prominent performance in capturing pairwise relationships. However, the heterogeneity and complexity of multi-modal medical data still pose significant challenges for standard GNNs, which struggle with learning higher-order, non-pairwise relationships. This paper proposes \textbf{GAMMA-PD} (\textbf{G}raph-based \textbf{A}nalysis of \textbf{M}ulti-modal \textbf{M}otor Impairment \textbf{A}ssessments in \textbf{P}arkinson's \textbf{D}isease), a novel heterogeneous hypergraph fusion framework for multi-modal clinical data analysis. \textbf{GAMMA-PD} integrates imaging and non-imaging data into a "hypernetwork" (patient population graph) by preserving higher-order information and similarity between patient profiles and symptom subtypes. We also design a feature-based attention-weighted mechanism to interpret feature-level contributions towards downstream decision tasks. We evaluate our approach with clinical data from the Parkinson's Progression Markers Initiative (PPMI) and a private dataset. We demonstrate gains in predicting motor impairment symptoms in Parkinson's disease. Our end-to-end framework also learns associations between subsets of patient characteristics to generate clinically relevant explanations for disease and symptom profiles. The source code is available at {\small \url{https://github.com/favour-nerrise/GAMMA-PD}}.
\end{abstract}

\section{Introduction}
Clinical decision support (CDS) systems rely heavily on heterogeneous data and their underlying biological, clinical, and statistical differences \cite{cios2002uniqueness}. Some examples include imaging, genomics, and electronic health records (EHRs). These diverse data types are highly interrelated and essential for informed clinical diagnosis, prognosis, and treatment. Traditional CDS systems, however, often fall short when integrating and interpreting voluminous and multifaceted information across different clinical data modalities. Consequently, there is a pressing need for robust and efficient methods that effectively leverage this data diversity and uncover deeper insights into disease mechanisms and patient care. 

An area largely affected by disease heterogeneity is the study of Parkinson's disease (PD), a neurodegenerative movement disorder that affects millions of people around the world \cite{shulman2011parkinson}. PD is characterized by a wide range of motor (e.g., bradykinesia, tremor, rigidity) and non-motor (e.g., cognitive decline, sleep disturbances, depression) symptoms that vary dramatically across patients in their onset, severity, and progression \cite{greenland2019clinical}. The Movement Disorder Society-Unified Parkinson's Disease Rating Scale (MDS-UPDRS) \cite{goetz2008movement} is a gold-standard clinical assessment for quantifying the heterogeneity of PD symptoms and progression. Conventionally, movement disorder specialists subtype PD into motor phenotypes such as postural instability and gait difficulty (PIGD) and tremor-dominant (TD) through extensive physical examinations and ratings using MDS-UPDRS. However, the assessments rely on subjective observations, which leads to inter-rater variability between clinicians. This challenge underscores the complexity and variability within the disease and the need for heterogeneous modalities and objective, quantitative biomarkers \cite{stebbins2013identify,mitchell2021emerging}. 

In response to the above challenges, computational methods for classifying PD symptoms have gradually shifted towards more sophisticated models, such as Graph Neural Networks (GNNs) \cite{velivckovic2017graph} that learn connections in structured data represented as graphs. These studies, however, often focus on single modalities (e.g., structural or functional neuroimaging) \cite{zhang2018multi,pan2021characterization,ji2022fc} where each brain region is a node in the graph. Some other works represent whole patient clinical data in a population graph where each patient is a node with features, and a semantic relationship like demographic similarity defines the edges between any two patients; however, they typically use homogeneous graphs (i.e., one node type), which only allow pairwise relationships between patients. Additionally, these works often use naive fusion methods such as clustering \cite{albrecht2022unraveling} for combining graph-based relationships and features from different data modalities, which makes it difficult to derive clinically meaningful interpretations that impact their study objective. Lastly, multimodal non-imaging and imaging studies predicting motor symptoms in PD \cite{leung2018using,adams2021improved} rely on Single Photon Emission Computed Tomography (SPECT) with the dopamine transporter-binding ligands (DaTSCAN) imaging or Position Emission Tomography \cite{raichle1983positron}, which is harder to acquire than resting-state fMRI (our imaging modality) and cannot capture functional abnormalities. 

To better derive clinically meaningful interpretations, this work creates a population graph in a heterogeneous hypergraph structure using imaging and non-imaging patient data to predict motor symptom progression in PD. Hypergraphs \cite{bretto2013hypergraph} capture patient-to-patient pairwise and patient-to-group higher-order relationships through hyperedges based on a single similarity. However, these structures fail to capture relationships between patients based on different similarity types (e.g., demographic and genomic similarities ) within a single graph \cite{liu2017multi}. Heterogeneous graphs extend this by modeling higher-order patient relationships through multiple edges (similarities) or nodes \cite{sun2021heterogeneous}. We introduce domain-specific hyperedge types based on medical knowledge clustering, enabling our model to learn clinically relevant patterns and relationships that are often overlooked in standard graph-based approaches. This approach is particularly suited to PD subtyping where patients who exhibit similar characteristics (present in non-imaging and imaging clinical data) are likely to be grouped within the same disease or symptom subtypes \cite{kim2023heterogeneous}.

\begin{figure}[t]
    \includegraphics[width=1\linewidth]{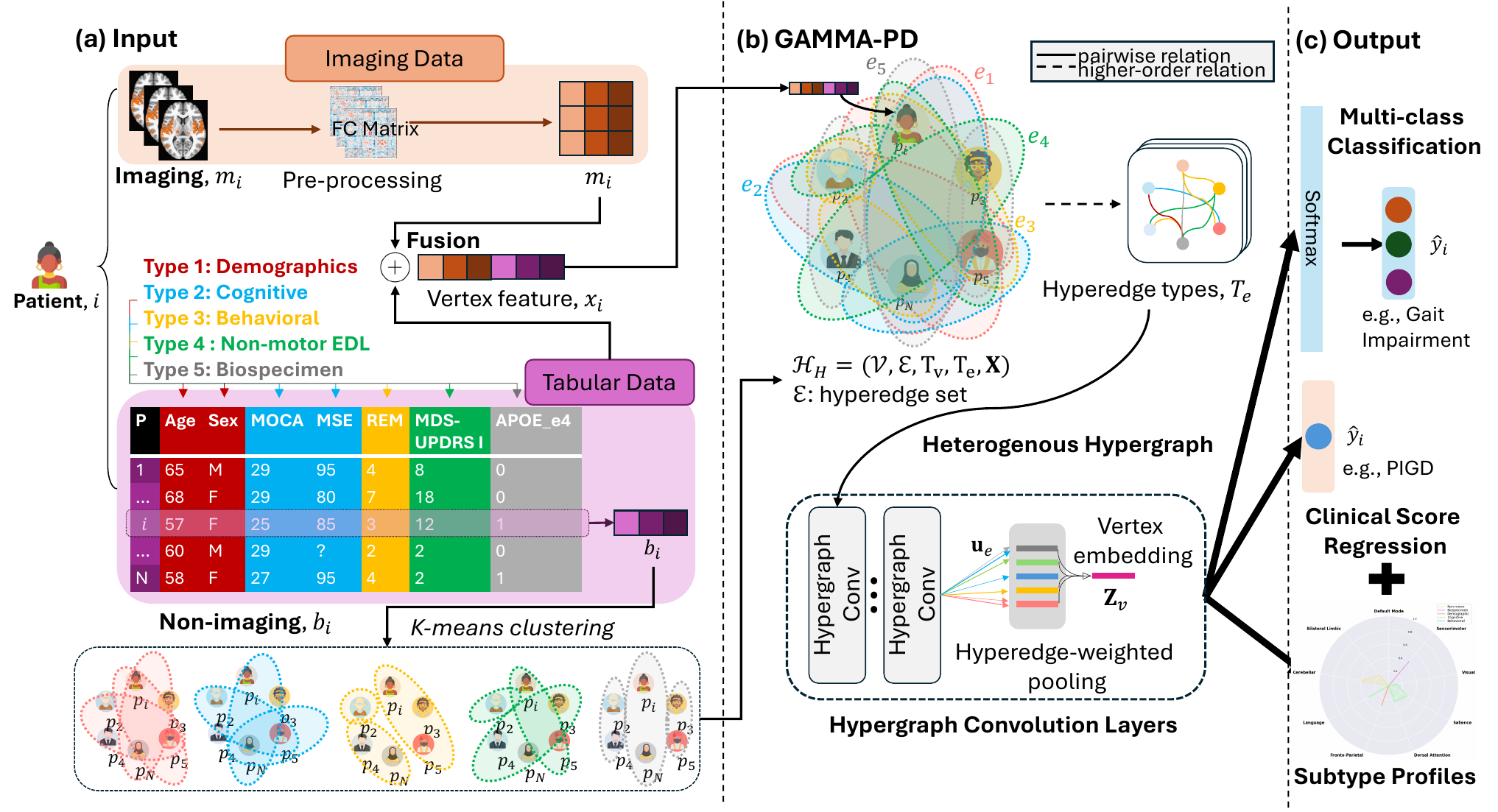}
    \caption{\textbf{\textbf{GAMMA-PD}}: (A) For each patient $i$, we pre-process their multi-modal medical record, which includes structured imaging data $m_i$ and unstructured non-imaging data $b_i$. We fuse both data into patient (node-level) features. We also cluster patient subgroups based on five similarity types from non-imaging data. (B) We form a population heterogeneous hypergraph containing all patients, their combined features, and relationships between them across different similarity types. This multiplex network is analyzed using neural networks to identify patterns within patient subgroups. (C) We classify the patients' gait impairment ratings (MDS-UPDRS 3.10), predict their TD/PIGD scores, and provide interpretable patient subtype profiles.}
    \label{fig:gamma-pd}
\end{figure}

Our proposed model, \textbf{GAMMA-PD} (Fig. \ref{fig:gamma-pd}), is a CDS-assistive tool for analyzing motor subtype heterogeneity in complex diseases like PD using multi-modal data, particularly neuroimaging, patient demographic characteristics, cognitive, behavioral, biospecimen, and non-motor experiences of daily living. Using \textbf{GAMMA-PD}, we cluster patients into various similarity subgroups using non-imaging features and train a heterogeneous hypergraph neural network to learn patient-to-patient and patient-to-group relationships for deriving distinct clinical subtype profiles. Additionally, we propose a custom feature-based attention-weighted message-passing mechanism to enhance the model's ability to identify and prioritize the most relevant features and relationships for each prediction task, providing a level of interpretability crucial for clinical applications. Accordingly, we aim to address the following challenges: (\textbf{a}) disease and symptom heterogeneity in neurodegenerative disorders like PD are complex to track and predict using a single modality type; (\textbf{b}) high-dimensional multi-modal medical data can obscure the most relevant feature interactions (\textbf{c}) CDS systems often lack domain knowledge integrated into model decision-making for clinical interpretability. To our knowledge, we are the first to propose a framework that uses heterogeneous hypergraph learning of multi-modal clinical data for improving the subtyping of PD symptoms and disease heterogeneity. 

In summary, we offer the following contributions:
(1) An end-to-end novel heterogeneous hypergraph pipeline \textbf{GAMMA-PD}, for learning higher-order interactions in complex, structured medical data. 
(2) A learnable feature-guided attention mechanism to enhance identification of relevant symptom interactions in high-dimensional data.
(3) A subtype profiler for interpreting associations between subtype imaging (particularly brain network) and non-imaging feature interactions.

\section{Methodology}
\textbf{Problem Definition} Given a multi-modal medical dataset $\mathcal{D} =\{{\mathbf{M},\mathbf{B},\mathbf{Y}}$, where $\mathbf{M} \in \mathbb{R}^{|\mathcal{N}|\times \mathcal{F}_{img}}$ is 2D or 3D medical imaging data, $\mathbf{B} \in \mathbb{R}^{|\mathcal{N}|\times \mathcal{F}_{non-img}}$ is non-imaging tabular medical data consisting of categorical and numerical features (demographics, cognitive variables, behavioral variables, non-motor progression, and biospecimen), $\mathbf{Y}={y_1, y_2, \ldots, y_{N}}$ is a set of ground-truth motor symptom severity measures, where $y_i \in \mathbb{R}$ for regression tasks or $y_i \in {1, \ldots, c}$ for classification tasks with $c$ discrete severity classes. Here, $N$ represents the number of patients, $\mathcal{F}_{img}$ is the dimensionality of imaging features, and $\mathcal{F}_{non-img}$ is the dimensionality of non-imaging features. Our objective is to predict the motor symptom severity $y_i$ for patient $i$ given their multi-modal medical data, i.e., medical image $\mathbf{m}_i \in \mathbf{M}$ and non-image data $\mathbf{b}_i \in \mathbf{B}$.

\noindent\textbf{Hypergraph}: A hypergraph is an extended graph $\mathcal{H}=(\mathcal{V},\mathcal{E}, \mathbf{W})$, where $\mathcal{V}=\{v_1, v_2, \ldots, v_N\}$ is a set of vertices (e.g., patients), $\mathcal{E}$ is the set of hyperedges (e.g., similarities between patients), and $\mathbf{W} \in \mathbb{R}^{|\mathcal{E}| \times |\mathcal{E}|}$ is a positive, diagonal hyperedge weight matrix, approximately equal to the identity matrix $\mathbf{I}$. Each hyperedge $e \in \mathcal{E}$ can connect multiple vertices simultaneously, allowing hypergraphs to model complex high-order relations beyond pairwise interactions. The relationship between vertices and hyperedges is represented by an incidence matrix $\mathbf{H} \in \mathbb{R}^{|\mathcal{V}|\times|\mathcal{E}|}$, where each non-zero entry $h(v,e)$ denotes $v \in e$. Additionally, each hypergraph contains a set of vertex features $\mathbf{X} \in \mathbb{R}^{|\mathcal{V}|\times d}$, where $\mathbf{x}_i$ denotes a vector of three-dimensional features $d$ for the vertex $i$.

\noindent\textbf{Heterogeneous Hypergraph}: A heterogeneous hypergraph $\mathcal{H}_H = (\mathcal{V}, \mathcal{E}, \mathcal{T}_v, \mathcal{T}_e, $ \\ $\mathbf{W})$ extends the concept of hypergraphs by incorporating multiple types of vertices $\mathcal{T}_v$ and hyperedges $\mathcal{T}_e$. This structure enables modeling complex relationships, such as types of similarities between patients. Let $\mathbf{D}_v \in \mathbb{R}^{|\mathcal{V}|\times|\mathcal{V}|}$ and $\mathbf{D}_e \in \mathbb{R}^{|\mathcal{E}|\times|\mathcal{E}|}$ be diagonal matrices containing vertex and hyperedge degrees, respectively. The hypergraph Laplacian is defined as $\Delta = \mathbf{I} - \mathbf{\Theta}$, where $\mathbf{\Theta} = \mathbf{D}^{-\frac{1}{2}}_v \mathbf{H}\mathbf{W}\mathbf{D}_e^{-1}\mathbf{H}^T \mathbf{D}^{-\frac{1}{2}}_v$ is the adjacency matrix. Our goal is to learn a representation $\mathbf{Z}^{\mathcal{H}_H} \in \mathbb{R}^{|\mathcal{V}| \times C}$, where each row represents the embedding of a vertex.

\subsection{Heterogeneous Hypergraph Construction}
To construct $\mathcal{H}_{H}$, we first obtain vertex feature embeddings $\mathbf{X}=\{\mathbf{X}_{1}, \mathbf{X}_{2}, \hdots, $ \\ $\mathbf{X}_{N}\}$ by concatenating pre-processed imaging features $\mathbf{M}$ and non-imaging features $\mathbf{B}$ (Fig. \ref{fig:gamma-pd}(a)). We pre-define five hyperedge types $\mathcal{T}e$ using medical domain knowledge: demographic, cognitive, behavioral, non-motor progression, and biospecimen. For each vertex's non-imaging feature vector $\mathbf{b}i$, we calculate cosine pairwise similarities between patients and group the similarity matrices into clusters via K-means \cite{abraham2014machine}. Each cluster forms a hyperedge where hyperedges belong to the same similarity type. Hyperedges are concatenated to form a final graph $\mathcal{H}_H$. We feed the hypergraph embeddings $\mathbf{X}^{\mathcal{H}_{H}}$ into Hypergraph Convolution Layers for additional analysis. 

\subsection{Hypergraph Convolution \& Attention Learning}
Hypergraph Neural Networks (HGNNs) \cite{bai2021hypergraph} are a powerful means to learn intricate dependencies in data by exploiting the rich, many-to-many relationships encoded in irregular data structures (i.e., structures beyond pairwise relations such as also triadic, tetradic or some higher-order relations). The convolution operation on a hypergraph $\mathcal{H}$ at layer $l$ is defined as follows:
\begin{equation}
    \mathbf{X}^{(l+1)} = \mathbf{\Theta}\mathbf{X}^{(l)}\mathbf{P}
\end{equation}
$\mathbf{P} \in \mathbb{R}^{F(l) \times F(l+1)}$ is the weight matrix between the $(l)$-th and $(l+1)$-th layer. 

\noindent\textbf{Feature-based Attention-weighted Message Passing}. We implement a custom feature-based attention message passing to enhance the model's ability to capture complex feature interactions. For each vertex $v$, we compute attention coefficients $\alpha_{v,e}$ for its incident hyperedges $e \in \mathcal{E}(v)$:
\begin{equation}
\alpha_{v,e} = \text{softmax}_{e \in \mathcal{E}(v)}(\text{LeakyReLU}(\mathbf{a}^T[\mathbf{P}\mathbf{x}_v | \mathbf{P}\bar{\mathbf{x}}_e]))
\end{equation}
where $\mathbf{a}$ is a learnable attention vector, $\mathbf{x}_{v}$ is the feature vector of vertex $v$, and $\bar{\mathbf{x}}_{e}$ is the mean feature vector of vertices in hyperedge $e$. The updated vertex features are then computed as:
\begin{equation}
\mathbf{x}_{v}^{(l+1)} = \sigma\left(\sum{e \in \mathcal{E}(v)} \alpha_{v,e}\mathbf{P}\bar{\mathbf{x}}_e\right)
\end{equation}
After propagating through $L$ layers, the weighted aggregated features for each $\mathbf{x}_{v}$ are passed through a fully connected layer with a non-linear activation function) to produce the final node embeddings: 
\begin{equation}
    \mathbf{Z}_v = \sigma (\mathbf{P}_f \cdot \mathbf{X}^{'}_v + \mathbf{u}_f),
\end{equation}
where $\mathbf{P}_f$ is the weight matrix of the final transformation layer, and $\mathbf{u}_f$ is the bias term.
The training objective is to minimize a loss function \(L(\theta)\), reflecting the discrepancy between the network's predictions \(\hat{y}\) and the true outcomes \(y\), parameterized by \(\theta\):
\begin{equation}
L(\theta) = \text{CrossEntropy}(\hat{y}, y) \quad \text{or} \quad L(\theta) = \text{MSE}(\hat{y}, y),
\end{equation}
depending on the task (e.g., classification or regression), where CrossEntropy and MSE denote cross-entropy loss and mean squared error, respectively.

\section{Experiments}
\textbf{Dataset}: We used two datasets: the Parkinson's Progression Markers Initiative (PPMI \cite{marek2018parkinson}) with 342 total participants (156 healthy controls and 186 participants with PD (mean age $64 \pm 9.7$ M=220, F=122)); a private dataset ~\cite{lu2021quantifying} of 35 participants with PD (mean age $69 \pm 7.9$) for holdout validation. PPMI inclusion criteria were: Healthy Control (HC) or PD participant, baseline visit, OFF medication, resting-state and 3D T1-weighted MRI brain imaging volumes present, biospecimen data available, and $\geq$50\% of non-imaging clinical records present. In total, we obtained 2,313 brain imaging scans to extract imaging features. We selected $28$ non-imaging clinical features (informed by a thorough systematic review \cite{leung2018using,albrecht2022unraveling,mestre2018reproducibility} See Supplementary Materials for additional information.

\textbf{Data Processing}: For missing non-imaging clinical features, we applied  Multiple Imputation with Chained Equation (MICE) \cite{azur2011multiple} for missing non-imaging data, and performed z-score normalization. Due to severe class imbalances for the severely impaired cohort, we binarized gait impairment scores to Mild (0, n=156) and Moderate (1-2, n=186). PIGD scores were calculated from MDS-UPDRS Parts II and III. We pre-processed MRI images using fMRIPrep and an AAL atlas parcellation \cite{tzourio2002automated} using nilearn \cite{nilearn,abraham2014machine}, resulting in $116 \times 116$ functional connectivity (FC) matrices where each entry in row $i$ and column $j$ in the matrix is the Pearson correlation between the average rs-fMRI signal measured in a brain region of interest (ROI) $i$ and ROI $j$. We z-score normalized the FC matrices and took the flattened upper right triangles as the imaging feature vectors.

\noindent\textbf{Setup}: Our model achieves its best performance with a 2-layer HypergraphConv (provided by PyTorch Geometric \cite{fey2019fast}), a hidden channel dimension of $64$, ReLU activation function, dropout rate of $0.2$, and $4$ multi-attention heads. We fine-tuned and selected the following hyperparameters on our PPMI validation set: a weighted AdamW optimizer \cite{loshchilov2017decoupled}, learning rate=$1e-3$, task loss regularization coefficient=$1e-2$, and $5$-fold cross-validation for $70$ epochs each.

\section{Results \& Discussion}
Table \ref{tab:perf_tab} presents a comprehensive evaluation on the holdout set for classification and regression tasks on predicting gait impairment severity and PIGD scores. We compare the performance of conventional methods, including MLP \cite{rumelhart1986learning}, GCN \cite{kipf2016semi}, GATv2 \cite{velickovic2017graph}, and BrainGNN \cite{li2021braingnn}, Linear Regression \cite{seber2012linear}, CPM \cite{shen2017using}, PNA \cite{corso2020principal}, RegGNN \cite{hanik2022predicting}, with our proposed method, \textbf{GAMMA-PD}, in various configurations. We assess \textbf{GAMMA-PD}'s performance over following ablation conditions: \textbf{GAMMA-PD} (M) with only imaging vertex features, \textbf{GAMMA-PD} (B) with only non-imaging vertex features, and \textbf{GAMMA-PD} with fused imaging and non-imaging vertex features, and \textbf{GAMMA-PD}+ with feature-based attention-weighted message passing.
\begin{table}[t!]
    \begin{subtable}[t]{.5\linewidth}
        \centering  
        \setlength{\tabcolsep}{1pt}
        \renewcommand{\arraystretch}{.75}
        \begin{tabular}{ lcccc } 
            \toprule
            \textbf{Method} & \textbf{Pre} & \textbf{Rec} & $\mathbf{F_{1}}$ & \textbf{AUC} \\
            \midrule
            MLP \cite{rumelhart1986learning} & 0.65 & 0.60 & 0.66 & 0.68\\
            GCN  \cite{kipf2016semi} & 0.64 & 0.66 & 0.65 & 0.66\\
            GATv2 \cite{velickovic2017graph} &  0.67 & 0.68 & 0.67 & 0.69\\
            BrainGNN \cite{li2021braingnn} & 0.69 & 0.71 & 0.71 & 0.73\\
            \midrule
            \textbf{GAMMA-PD} (M) & 0.72 & 0.74 & 0.73 & 0.76\\
            \textbf{GAMMA-PD} (B) & 0.73 & 0.75 & 0.74 & 0.78 \\
            \midrule
            \textbf{GAMMA-PD}* & 0.75 & 0.79 & 0.77 & 0.81 \\
            \rowcolor{Gray}
            \textbf{GAMMA-PD}+* & \textbf{0.81} & \textbf{0.81} & \textbf{0.83} & \textbf{0.86} \\
            \bottomrule
        \end{tabular}
        \caption{Classification Results - Gait}
        \label{tab:classification_res}
    \end{subtable}%
    \quad
    \begin{subtable}[t]{.5\linewidth}
        \centering
        \setlength{\tabcolsep}{1pt}
        \renewcommand{\arraystretch}{.75}
        \begin{tabular}{ lcccc } 
            \toprule
            \textbf{Method} & \textbf{RMSE} & \textbf{MAE} \\
            \midrule
            Linear \cite{seber2012linear} & 7.80 & 5.24 \\
            CPM \cite{shen2017using} & 7.52 & 4.56 \\
            PNA \cite{corso2020principal} & 7.71 & 4.60 \\
            RegGNN \cite{hanik2022predicting} & 7.66 & 5.53 \\
            \midrule
            \textbf{GAMMA-PD} (M) & 6.38 & 3.29 \\
            \textbf{GAMMA-PD} (B) & 6.44 & 3.62 \\
            \midrule 
            \textbf{GAMMA-PD}* & 5.75 & 3.51 \\
            \rowcolor{Gray}
            \textbf{GAMMA-PD}+* & \textbf{2.31} & \textbf{1.89} \\
            \bottomrule
        \end{tabular}
        \caption{Regression Results - PIGD Score}
        \label{tab:regression_res}
    \end{subtable}
    \caption{(a)  MDS-UPDRS Part 3.10 gait impairment severity [0 (Mild), 1 (Moderate)]; (b) MDS-UPDRS PIGD Score [0,0.80]. (M) imaging features only; (B) non-imaging features only; (+) fused features w/ feature-based attention-weighted pooling; * denotes statistical significance ($p < 0.05$, Wilcoxon signed rank test) compared with our best-performing results.}
    \label{tab:perf_tab} 
\end{table}

For the classification task (Table \ref{tab:classification_res}), the reported metrics are macro-weighted Precision (Pre), Recall (Rec), $F_1$-score, and the Area Under the ROC Curve (AUC). \textbf{GAMMA-PD}, containing fused imaging and non-imaging features, outperforms all other methods with significant differences ($p \leq 0.05$), achieving an $F_1$score of 0.75 and an AUC of 0.79, as determined by the Wilcoxon signed-rank test. Notably, \textbf{GAMMA-PD}+, enabled with feature-based attention-weighted message passing, enhances the performance and reaches the highest $F_1$-score of 0.81 and an AUC of 0.84. 

For regression tasks predicting the PIGD score (Table \ref{tab:regression_res}), the reported metrics are the Root Mean Square Error (RMSE) and Mean Absolute Error (MAE). The \textbf{GAMMA-PD} domain-driven hyperedge type (dm) shows notable improvements over traditional methods like Linear Regression \cite{schneider2010linear} and PNA \cite{corso2020principal}. \textbf{GAMMA-PD}+ again demonstrates superior performance with the lowest RMSE of 14.31 and MAE of 10.89, significantly outperforming all other methods. These results indicate the robustness of the \textbf{GAMMA-PD} model, particularly in its \textbf{GAMMA-PD}+ configuration, for both predictive tasks related to gait disturbances. The classification and regression results for all metrics suggest that \textbf{GAMMA-PD}, especially when using heterogeneous fused data modalities, can capture the complex patterns underlying motor impairments in PD more effectively than existing baselines.

\section{Clinical Interpretability}
\begin{figure}[t]
    \begin{minipage}[b]{0.43\textwidth}
        \includegraphics[width=1\linewidth]{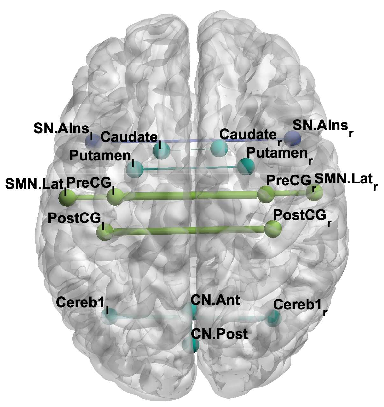}
        \caption{Top-10 thresholded ROI connections via attention-based brain networks for predicting PIGD score SensoriMotor Network, Salience Network, and Cerebellar Network.}
        \label{fig:glass_brains}
    \end{minipage}~
    \begin{minipage}[b]{0.55\textwidth}
        \includegraphics[width=1\linewidth]{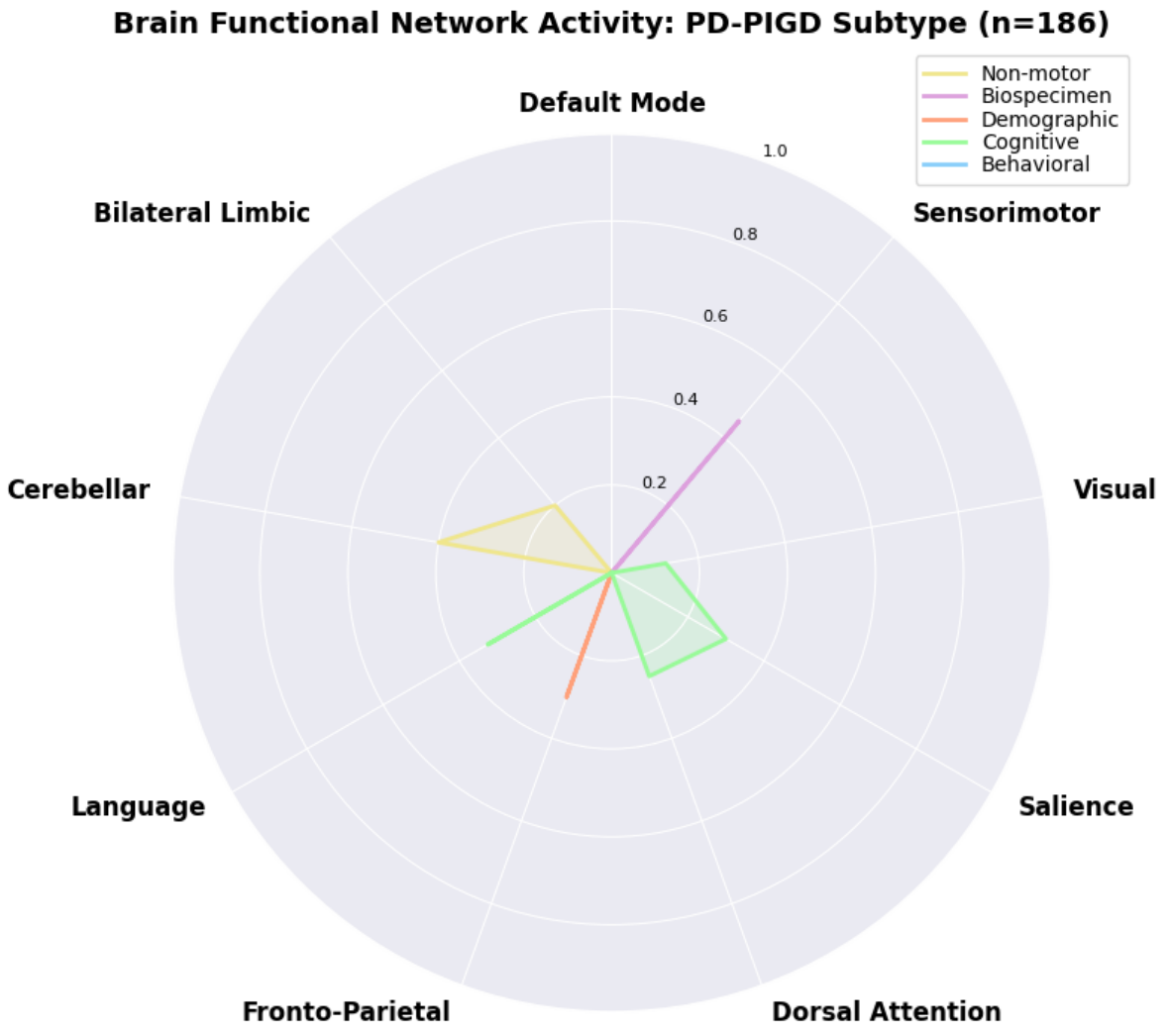} \vspace{-20pt}
        \caption{Subtype profile of PD-PIGD predictions characterized by feature similarity activity within brain functional networks.}
        \label{fig:radial}
    \end{minipage}
\end{figure}

For clinical interpretability of the ROIs most associated with predicting the PIGD subtype, we first generated brain network visualizations (shown in  \cref{fig:glass_brains}) by partitioning the ROIs into nine ``networks based on their functional roles: Default Mode Network (DMN), SensoriMotor Network (SMN), Visual Network (VN), Salience Network (SN), Dorsal Attention Network (DAN), Fronto-Parietal Network (FPN), Language Network (LN), Cerebellar Network (CN), and Bilateral Limbic Network (BLN), denoted by nodes. The edge widths here are the attention weights between networks. Overall, we observe several interactions within CN, particularly Crus I, which is significantly associated with overt motor and sensorimotor movement and planning \cite{hou2018patterns,ma2017disrupted}. We also see connections between the precentral gyrus (frontal lobe) and the postcentral gyrus (parietal lobe), which play an important role in voluntary movement and are associated with postural deformities and gait freezing \cite{teramoto2014relation}. Lesser but significant connections also exist within the cerebellar regions of the CN located in the basal ganglia (e.g. Putamen and Caudate), which are known to have dysfunction in PD due to striatal dopamine deficiency \cite{yang2013changes,wu2009regional}. 

\cref{fig:radial} shows brain functional network activity patterns in patients profiled with a PD-PIGD subtype. It shows a predominant activation of motor-related regions within the SMN (particularly in the Supplementary Motor Area, Primary Motor Cortex, and Basal Ganglia regions), which reflects substantial changes in motor-related activity (herein interpreted as a decrease compared to the non-PID-PIGD case). This aligns with previous findings where these changes may be understood as a result of decreased striatal activations following dopaminergic denervation \cite{gilat2019functional}. The CN (Cerebellum region) also showed moderate activity and has shown to be significantly associated with lower overt motor and sensorimotor movement and planning in a gait-impaired PD state \cite{hou2018patterns,ma2017disrupted}. Additionally, cognitive-linked networks like the FPN (Prefrontal Cortex region) and SN (Anterior Cingulate Cortex and Insula regions) are also highlighted, supported by previous research such as \cite{teramoto2014relation,mi2017altered,wolters2019resting} that demonstrate these networks play a strong role in voluntary movement and are associated with postural deformities and freezing of gait. These findings emphasize the complex interplay between motor, cognitive, and visual processes in PD motor impairment subtypes. This multifaceted nature underscores potential targets for therapeutic interventions to enhance gait function in PD patients.

\section{Conclusion}
We propose \textbf{GAMMA-PD}, a hypernetwork fusion framework for multi-modal imaging and non-imaging analysis of motor impairment subtyping in Parkinson's Disease. Rigorous evaluations on several PD medical datasets showed the advantage of our method in assessing gait impairment severity in PD and postural instability subtype differences. We also constructed clinically meaningful interpretations to contextualize which features mostly strongly contributed to model predictions and predicted output correlates to brain functional networks. Future works include expanding our dataset sources, extending our analysis to a longitudinal study to capture symptom progression, investigating additional baselines and network learning mechanisms, and further validating clinical interpretations and ethical implications. 
\\
\\
{\small \noindent\textbf{Acknowledgements.} This work was partially supported by NIH grants (AA010723, NS115114, P30AG066515), the Stanford School of Medicine Department of Psychiatry and Behavioral Sciences Jaswa Innovator Award, the Stanford Institute for Human-Centered AI (HAI) Google Cloud credits, and HAI Hoffman-Yee Award. FN is funded by the Stanford Graduate Fellowship, the Stanford NeuroTech Training Program Fellowship, the Stanford HAI Graduate Fellowship, and the Stanford McCoy Family Center for Ethics in Society Graduate Fellowship.} 

%
\bibliographystyle{splncs04}
\bibliography{paper0016}

\end{document}


\title{Supplementary Material for ``GAMMA-PD: Graph-based Analysis of Multi-Modal Motor Impairment Assessments in Parkinson's Disease"}
\author{}
\institute{}
\maketitle    
\section*{Non-imaging, clinical features from the Parkinson's Progression Markers Initiative (PPMI) Dataset}
We extract 28 total non-imaging, clinical features and 4 motor-related features (ground-truth labels) from the PPMI dataset and group them into similarity types:
\begin{itemize}
    \item \textbf{Demographic features}: sex (Male=1, Female=0), years of education (1=$<$13 2=13,23], 3=23+, NaN=Unknown), race (1=White, 2=Black 3=Asian, \\
    4=Other), family history of PD (1=yes, 2=no), handedness (1=right, 2=left, 3=mixed).
    \item \textbf{Cognitive features}: Montreal Cognitive Assessment (MoCA) score \cite{nasreddine2005montreal} adjusted for education, Modified Schwab and England Capacity for Daily Living Scale (S\&E ADL Scale) Score \cite{schwab1969projection}, Geriatric Depression Scale (GDS) score \cite{yesavage1982development}.
    \item \textbf{Behavioral features}: Scales for Outcomes in PD-Autonomic (SCOPA-AUT) Score \cite{marinus2004short}, REM Sleep Behavior Disorder Questionnaire Score \cite{stiasny2007rem}, UPSIT Age-/Sex-Adjusted Percentile, Epworth Sleepiness Scale Score (ESS) \cite{johns1991new}, Questionnaire for Impulsive-Compulsive Disorders in Parkinson's Disease–Rating Scale (QUIP-RS) Score \cite{weintraub2009validation}.
    \item \textbf{Biospecimen features}: Number of E4 alleles in apolipoprotein E (APOE) genotype, Cerebrospinal fluid (CSF) A-beta 1-42, CSF Alpha-synuclein, CSF t-tau, CSF p-tau, CSF Neurofilament Light, and Serum Uric Acid (mg/dL). 
    \item \textbf{Non-motor experiences of daily living}: MDS-UPDRS Part I Anxious Mood, Apathy, Cognitive Impairment, Dopamine Dysregulation Syndrome, Depressed Mood, Fatigue, Hallucinations and Psychosis scores \cite{goetz2008movement}.
    \item \textbf{Motor features}: MDSUPDRS Part III.10 Gait Impairment score, PIGD dominant score (Mean of MDS-UPDRS NP2WALK, NP2FREZ, NP3GAIT, NP3FRZGT, NP3PSTBL ratings taken when participant is either "OFF" PD medication or untreated for PD), Tremor dominant score (mean MDS-UPDRS NP2TRMR, NP3PTRMR, NP3PTRML, NP3KTRMR, NP3KTRML, \\
    NP3RTARU, NP3RTALU, NP3RTARL, NP3RTALL, NP3RTALJ, \\
    NP3RTCON ratings taken when participant is either "OFF" PD medication or untreated for PD), and TD/PIGD classification (Tremor score / PIGD score; if ratio $\geq$1.15, OR if PIGD score = 0 and Tremor score $>$ 0, then participant is TD.  If ratio $\leq$ 0.9 then participant is PIGD).  
\end{itemize}
